\documentclass{article}

\usepackage{microtype}
\usepackage{graphicx}
\usepackage{subfigure}
\usepackage{booktabs} 

\usepackage{hyperref}

\usepackage[accepted]{hill2019}

\icmltitlerunning{Crowdsourcing in the Absence of Ground Truth---A Case Study}

\begin{document}

\twocolumn[
\icmltitle{Crowdsourcing in the Absence of Ground Truth---A Case Study}



\icmlsetsymbol{equal}{*}

\begin{icmlauthorlist}
\icmlauthor{Ramya Srinivasan}{fla}
\icmlauthor{Ajay Chander}{fla}
\end{icmlauthorlist}

\icmlaffiliation{fla}{Fujitsu Laboratories of America, Sunnyvale, CA, USA}

\icmlcorrespondingauthor{First Author}{ramya@us.fujitsu.com}

\icmlkeywords{Crowdsourcing, experienced emotions, subjective opinions}

\vskip 0.3in
]



\printAffiliationsAndNotice{ElicC} 

\begin{abstract}
Crowdsourcing information constitutes an important aspect of human-in-the-loop learning for researchers across multiple disciplines such as AI, HCI, and social science. While using crowdsourced data for subjective tasks is not new, eliciting useful insights from such data remains challenging due to a variety of factors such as difficulty of the task, personal prejudices of the human evaluators, lack of question clarity, etc. In this paper, we consider one such subjective evaluation task, namely that of estimating experienced emotions of distressed individuals who are conversing with a human listener in an online coaching platform. We explore strategies to aggregate the evaluators choices, and show that a simple voting consensus is as effective as an optimum aggregation method for the task considered. Intrigued by how an objective assessment would compare to the subjective evaluation of evaluators, we also designed a machine learning algorithm to perform the same task. Interestingly, we observed a machine learning algorithm that is not explicitly modeled to characterize evaluators' subjectivity  is as reliable as the human evaluation in terms of assessing the most dominant experienced emotions. 
\end{abstract}

\section{Introduction}
Crowdsourcing is common in human-in-the-loop learning systems wherein data for a task is obtained through the services of a large number of people. This data could be used for training a machine learning algorithm, or could be used independently to make various decisions. 

Across a wide range of sectors such as marketing, advertising or industrial design, crowdsourcing has made a significant impact. Crowdsourcing is also making its way into more critical fields such as healthcare \cite{swan}, and is proving to be a faster alternative than conventional methods for predicting the spread of infectious diseases \cite{churana}, for diagnosis and treatment \cite{meyer}, and other healthcare applications.

In machine learning systems, crowdsourcing can aid in several aspects such as in: producing data, debugging and checking of models \cite{wu}, for active learning \cite{yan}, and to improve human-computer interaction in multi-agent systems \cite{abhigna}. 

Data creation, perhaps, remains one of the most common purposes of crowdsourcing. This includes using humans-in-the-loop for curating the data, for pre-processing and cleaning the data, and for generating labels. In most cases, generating labels is a straightforward task (e.g., object classification, face recognition, parts of speech tagging, etc.). However, in some applications (e.g., evaluating aesthetics of an image, assessing quality of machine-generated music, etc.), there could be ambiguity about the ground truth due to the subjective nature of the task. In such applications, eliciting ground truth from noisy human evaluations becomes a challenge. 

There has been an active line of research focusing on addressing the aforementioned challenge of eliciting ground truth in highly subjective tasks \cite{anca,felt,subramanian,giancola,procaccia}. Strategies vary from a simple majority voting consensus to more sophisticated techniques such as multi-annotator statistical models, and prior knowledge models. Motivated by these works, we consider a subjective task --- that of assessing {\it experienced emotions} --- to compare the performance of a simple voting consensus scheme with an optimal aggregation methodology \cite{procaccia}. This task offers an excellent scenario to study some challenges associated with human evaluations and to analyze how human evaluation compares to assessments from machine learning systems. Specifically, our problem setting and research questions are as described next.

\subsection{Setting}
We consider a conversational user interface (CUI) that is designed to support textual conversations between people needing emotional support (i.e. the users) and trained counselors who offer listening and support on the backend of the conversation platform (i.e. the human listeners). First, depending on the type of issue discussed between the user and the human listener, the conversation could involve varying amounts of emotional content.  Some of these emotions might be explicitly expressed in the conversation while others are only felt or experienced within the user \cite{ochs}. For example, a user may not be in touch with feeling scared and may instead express anger. Knowing the experienced emotions of a user can help in understanding their internal states and in addressing their concerns better. In short, the problem we consider is the assessment of experienced emotions of emotionally distressed users based on their textual conversations with human listeners. 

We deliberately choose experienced emotion assessment as this is more subjective than expressed emotion assessment. This problem is a highly subjective task. Each evaluator judges the user’s experienced emotion based on their personal experiences, socio-cultural backgrounds, and introspective abilities \cite{mcarthy}. Furthermore, collecting information about their experienced emotions from the distressed users themselves is highly unreliable due to the following reasons:
\begin{itemize}
{\item {\it Limited introspective capabilities of users:} A recent study involving over 800 studies of self-awareness indicated that emotionally distressed people have limited self-introspection abilities, and response biases \cite{eurich}. }
\item {{\it Transient nature of problems:} Many emotional conditions are short lived in terms of their time duration, intensity, frequency of occurrence, and often go unnoticed \cite{morris}. People with such problems appear normal in most ways and thus it is hard to even recognize that they have a problem. }
\item { {\it Social stigma:} People are often less comfortable in opening up about such problems as they fear a social stigma associated with the treatment \cite{ahmedani}}
\end{itemize}

Thus, experienced emotion assessment is a highly subjective task with no objective ground truth available. 

\subsection{Research Questions (RQs)}
The broad goal of our study is to explore if crowdsourcing is always helpful in extracting reliable information in highly subjective scenarios. Specifically, does crowdsourcing clarify or confound for highly subjective tasks? We explore this question by considering the problem setting described earlier. Within that setting, we investigated the following questions:
\begin{itemize}
    \item {RQ1: Are some optimum evaluator aggregation strategies such as \cite{procaccia} better than simple majority voting consensus for highly subjective tasks such as for the one considered?. }
\item {RQ2: For the task of experienced emotion assessment, how does a machine-learning algorithm that is not explicitly modeled to handle evaluator subjectivities compare with some common crowdsourcing aggregation strategies?}

\end{itemize}

\subsection{ Key Findings}
We list some findings below. A detailed description can be found in the Section 5. 
\begin{itemize}
    \item 
{For the task considered, a simple voting consensus scheme is more or less  as effective as an optimal aggregation strategy. }
\item {For the task of choosing the top experienced emotion, the machine’s result matched the human evaluators results for 75\% of the instances. For the task of choosing the top 3 experienced emotions (i.e. rankings) among the 6 emotions, the machine’s evaluation matched the human evaluation for roughly 50\% of instances.}

\end{itemize}

\section{Related Works}
We review related work pertaining to crowdsourcing for emotional/mental healthcare and crowdsourcing for subjective tasks. 
\subsection{Crowdsourcing for Mental Healthcare}
Mental health problems have become very common globally. A recent study reported that in the United States alone, about 56\% of adults with mental health conditions do not receive the treatment they need \cite{nguyen}. This, in turn, has triggered an interest in developing crowdsourcing platforms for treating mental health conditions.
Yet, there are only a few works that look at crowdsourcing techniques for addressing mental health conditions. 

In \cite{morris}, the author presents an online intervention called Panoply that administers emotion-regulatory support. In another work by \cite{naslund}, the authors surveyed the effect of randomized trials using online crowdsourced methods for recruitment, intervention delivery and data collection in people with mental conditions such as schizophrenia. As can be noted from these illustrations, most crowdsourcing platforms for healthcare focus on physiological conditions or well-defined mental conditions. In this work, we study the performance of crowdsourced evaluations for the assessment of emotional health conditions.

\subsection{Crowdsourcing for Subjective Tasks}
It has long been established that crowdsourcing methods can offer a reasonably accurate solution for subjective tasks. In \cite{rainer}, the authors analyze the benefit of using crowdsourcing for estimating the media playout in a multimedia system. In \cite{ghadirayam}, the authors evaluate the performance of crowdsourced data for assessing picture quality. In \cite{hobfeld}, the authors survey methods for assessing the quality of crowdsourced data for multimedia quality of experience.   The authors in \cite{checco} propose an elegant technique to aggregate individual responses for interval data. However, this method is not applicable to nominal or ordinal scales, which is the focus of this study. The authors of \cite{alfaro} propose a method wherein the evaluators are asked to compare items to obtain top k lists. In a similar vein, the authors in \cite{procaccia}, propose a pairwise comparisons followed by estimating a minimum feedback arc set of the tournament to optimally aggregate the uncertain preferences. We evaluate the feasibility of such methods in our study and show that pairwise comparisons are not very efficient for the problem considered. 

\section{Data Collection}
The dataset consists of four distinct user conversations with a single human listener (HL), and was collected using a CUI focused on connecting users with HLs. The total duration of the conversations was over two hours. The participants consented to the use of anonymized conversations for research and presentation purposes. Conversations between users and HLs dealt with topics such as relationship issues, work stress, etc.  Each conversation was divided into transcripts. A transcript is defined as a part of the conversation wherein the user is continuously engaged in ex-pressing themselves for more than three minutes at a stretch, independent of any interleaving HL responses.  

As such conversations are mostly about personal issues, user privacy preferences constrain data collection. Not all users are comfortable sharing their data.  This is a major bottleneck in being able to accumulate larger datasets of this nature that involve sensitive personal data. Many users abruptly left the CUI as the conversations progressed to-wards their personal issues. Additionally, not all transcripts specifically deal with emotionality; some transcripts consisted of the user and HL getting to know each other, engaging in back and forth small talk before any actual issues surfaced. Thus, although there were over fifty transcripts and 16 users in the dataset, emotion-related content was only available for twelve transcripts across 4 different users. We report results on this subset.

\vspace{0.2in}
{\bf Illustrative Transcript:} \\
{\it User:} Hi, can you please help me with anxiety. \\
{\it HL:} I'm sorry you're feeling anxious. Can you tell me more about it? \\
{\it User:} I have no self-confidence and have a girlfriend who I really like. I can't cope thinking she is going to find someone better. I am drinking to kill the anxiety. \\
{\it HL:} It sounds like you're feeling really anxious about your girlfriend staying with you. That sounds really difficult.\\
{\it User:} She is out with work tonight and a colleague who she dated for a bit is there. I don't know how to cope. \\
{\it HL:} It sounds like you're feeling really anxious that she is out with other people including her ex. And you not being there with her is making you feel worse. I'm sorry - that's a really hard feeling. \\
{\it User:} Can you help? \\
{\it HL:} I can listen to you. And I really am sorry that you're feeling so anxious. Maybe you can tell me more about your relationship and why you are feeling insecure.\\ 
{\it User:} I am an insecure person. I am a good-looking guy, always get chatted up, but I have no confidence.\\

\section{Amazon Mechanical Turk Survey}
We conducted a survey on Amazon Mechanical Turk (AMT) to investigate how human evaluators assessed experienced emotions in the conversation snippets. All 195 participants were US residents, about 62\% were male and 38\% female. Eighty one percent of them had attended college. Seventy percent of them were aged between 25 and 49. 
   
First, participants were asked a screening question regarding prior active listening experiences (i.e. participation in any of the following relevant fields: counseling, psychology, psychiatry, nursing, caregiving, non-violent communication class, active listening training, mediation). We eliminated 79 people who did not have any relevant experience. 

In order to ensure that the participants had basic active listening abilities, we conducted another screen. Participants were given three conversation snippets and were instructed that each conversation snippet was an excerpt of a longer conversation between someone who was seeking counseling for their problems (i.e. a user) and someone who had offered to listen to their problems (i.e. a human listener). Note that these three conversation snippets did not correspond to the conversations in our dataset, but were designed for screening. They were asked to read each conversation snippet and answer the question: What is the primary emotion that the user is expressing in this conversation snippet? They were presented with six options: Angry, Happy, Sad, Scared, Surprised, and Worried, and asked to select one.  For these three test cases, the answers were designed to be relatively easy and therefore seven participants who answered incorrectly were excluded from the study.
  
The final part of the survey consisted of presenting actual transcripts from our dataset. They were asked to read each transcript and answer the following question, with the associated guidance:

{\it Which emotions might the user be experiencing in this transcript text? To answer this, we would like you to look beyond the text and infer what you think the user might be experiencing. It can be a little tricky to see how an experienced emotion is different from an expressed emotion so here is a hypothetical example --- in some cultures, some people may not be comfortable directly expressing anger so they might express sadness instead.} 
   
First, participants were asked to choose the top experienced emotion from the 6 emotion choices mentioned earlier. Second, they were also asked to rank the top 3 experienced emotions given in the order of their likelihoods (from most likely experienced to least likely experienced). We did not ask the Turkers to rank all 6 because of the cognitive burden it imposes on the Turker (which, in turn, would lead to the increased probability of providing noisy assessments). That is, ranking all 6 emotion choices would be too difficult a task. For example, if surprised or happy are never mentioned or implied in the text, it is not possible to rank them either. 

\section{ Analysis of Research Questions}

We analyze the results of the AMT survey in the context of the research questions (RQs) described earlier. 


\subsection{RQ1: Simple Voting Consensus vs Optimal Aggregation}
The goal here is to compare the performance of an optimum voting aggregation strategy \cite{procaccia} with that of a simple majority voting consensus. We provide an overview of the method employed, the results obtained, and discuss the implications. 
\subsubsection{Method}
While many methods have been proposed to effectively aggregate subjective evaluations, they are for specific tasks as outlined in Section 2.2. The closest pertinent one for our purposes is the method suggested in \cite{procaccia} wherein the authors propose to extract reliable responses through ranking preferences. We wish to evaluate the efficacy of this method for the task of estimating experienced emotions. We compare this method with a simple voting consensus scheme wherein the emotion with the highest number of votes/preferences is considered to be representative of the most experienced emotion. As is evident, the method suggested in \cite{procaccia} is only applicable for the case of ordinal data. 

An uncertain evaluator response can be viewed as a distribution over rankings. A confident evaluator will report a single emotion whereas a highly uncertain evaluator will report all emotions. As elaborated in \cite{procaccia}, this can be formulated as the popular NP-hard problem of finding the minimum feedback arc set of a tournament.  

Specifically, the set of possible experienced emotions constitute the vertices of a directed graph. The weights $w_{ab}$ of this directed graph are determined by the number of evaluators preferring emotion a to emotion b as the experienced emotion. This directed graph with two weighted edges between each pair of vertices (one in each direction) is called a weighted tournament. The minimum feedback arc set of a tournament (also known as minimum feedback ranking) is the problem of finding the ranking of the vertices such that the sum of weights of edges that disagree with the ranking provided by the evaluators is minimized. In other words, this is same as the popular voting rule called Kemeny rule \cite{lv} which finds a ranking that minimizes the sum of Kendall Tau (KT) distances from the input rankings. The authors in \cite{procaccia} show that the minimum feedback ranking of a tournament with weights defined in this manner minimizes the expected sum of Kendall Tau distances from the evaluator preferences. 

The method suggested in \cite{procaccia} requires the weights $w_{ab}$ for all possible emotions. It is to be noted that the evaluators only ranked the top 3 experienced emotions. However, without loss of generality, the emotions not chosen by an evaluator are considered to have a lower preference/ranking for that evaluator. Also, in obtaining individual evaluator rankings across all the 6 emotions under consideration, we assume equal ranks for all the emotions not ranked/preferred by an individual evaluator (since the evaluator is asked to pick 3 out of the 6 choices).
   
We use a simple voting consensus method for comparison with the aforementioned method. The higher number of votes an emotion garners, the higher is its ranking. In this manner, the top 3 emotions are determined. 

\subsubsection{Results and Discussion}
Surprisingly, for the aforementioned task, the top 3 emotions as determined by the optimal aggregation method was the same as that obtained by the simple voting consensus method. 

The above result indicates that in highly subjective scenarios, a simple voting consensus is as effective as a ranking scheme (if not better). This is somewhat intuitive. When someone is highly uncertain, their uncertainty increases when they are asked to perform additional tasks. Selecting an emotion is easier than selecting and ranking them.  So, the extra task of ranking brings in additional uncertainty. 

\subsection{RQ2: Performance of a machine learning algorithm that is uninformed about human subjectivities}
The goal here is to compare the performance of an objective assessment from a machine learning algorithm (i.e. the algorithm is uninformed about human prejudices) in the subjective evaluation task under study. 
We designed a machine learning algorithm without explicitly modeling the subjective prejudices of the human evaluators. As a reference, a brief summary of the algorithm is provided here. 

\subsubsection{Method}
Motivated by the fact that Bayesian methods have been successful in modeling several aspects of human cognition \cite{grifthis}, we propose a novel Bayesian framework that fuses information about expressed emotion probabilities (which may be computed from existing emo-tion recognition methods) and sentiment embeddings \cite{tang} to compute probabilities of experienced emotions specific to individual users. 

We represent the probability of an emotion $i$ being experienced as $P(emo_{experienced}=i)$. An emotion recognition algorithm is run on the user's texts to determine the probability of all expressed emotions. Specifically, we construct a dictionary of synsets or synonyms for an emotion and based on the number of times a word appears in the conversation, a probability of an expressed emotion is calculated. We represent the probability of an emotion $j$ being expressed as $P(emo_{expressed}=j)$. 

Next, we use sentiment embeddings \cite{tang} to measure similarities between words depicting a pair of emotions. This information is computed across several people and through large datasets that contains emotional content (e.g. blogs, news articles, etc.). This is thus reflective of the general relatedness between two emotion-indicating words. These similarity measures are then normalized. Specifically, this constitutes the likelihood probability of expressing one emotion given that another emotion is experienced and is denoted by $P_l(emo_{expressed}=j | emo_{experienced}=i)$. Let $r_{emo-i,emo-j}$ be the measure of relatedness between two emotions ($i and j$) \cite{tang}. We compute likelihood probability $P_l(emo_{expressed}=j | emo_{experienced}=i)$ based on normalizing the similarities $r_{emo-i,emo-j}$  over the space of all possible emotions that are observed as 
\begin{equation}
    P_l(emo_{expressed}=j | emo_{experienced}=i)=  R
\end{equation}

where $R=P_l(emo_{expressed }=j|emo_{experienced}=i)$.

Using the Bayes rule of total probability, the probability of expressing an emotion $j$, i.e., $P(emo_{expressed}=j)$ can be written as 
\begin{equation}
P(emo_{expressed}=j)=\sum_{emo_{experienced}} (R*Q)
\end{equation}
wherein $Q$ is defined as follows: \\ \\\hfill $Q = P(emo_{experienced}=i)$. \\ \\
Note that the experienced emotion probabilities, i.e.,  $P(emo_{experienced}=i)$ for all possible emotion state variables are of interest to us. 
Without loss of generality, let us assume that there are $m$ possible expressed emotion states and $n$ possible experienced emotion states and that $n>m$ (Our framework applies to cases where 
$n<=m$ as well).
If the probabilities of all the experienced emotions of interest are denoted by the $n$ dimensional column vector $\bf{x}$, the likelihood probabilities for all possible $m.n$ emotion pairs by the matrix $\bf{L}$, and the probabilities of all expressed emotions by the $m$ dimensional column vector $\bf{t}$, then the resulting system of equations can be written as a constrained optimization problem as follows:
\begin{equation}
\min ||\bf{Lx-t}||_2^2
\end{equation} subject to the constraints $||\bf{x}||_1=1$ and $x_i>=0$, for all $i$.
The aforementioned equation can be solved by applying Karush-Kuhn-Tucker (KKT) conditions \cite{kkt} as follows:
\begin{equation}
D||{\bf Lx-t}||_2^2 +\lambda D({\bf x^T 1} -1) +{\bf \mu x}=0
\end{equation}
In eq. (4), { \bf1} is a $n$ dimensional column vector of 1, $\lambda$ is the Lagrange multiplier, ${\bf \mu} = [\mu _{1}, \mu_{2}, ....\mu_{n}] $ is the KKT multiplier such that $\mu {\bf x}=0 $ and $\mu_i$ $<$ 0 . $D$ denotes the derivative. 

\subsubsection{Results and Discussion}
For the task of choosing the top experienced emotion, the machine’s result matched the human evaluators results for 75\% of the transcripts. For the task of choosing the top 3 experienced emotions (i.e. rankings) among the 6 emotions, the machine’s evaluation matched the human evaluation for roughly 50\% of transcripts. 

As is evident from the algorithm, the machine provides probabilities for possible experienced emotions. The set of probable experienced emotions chosen by the machine was always lesser than the set of probable emotions chosen by the human evaluators. As a result, the uncertainty associated with the machine evaluation was lesser than the uncertainty associated with human evaluation. 

The machine’s assessment is based on the training data. As described in Section 5.3.1, likelihood probabilities are computed using large volumes of public text corpus. Such corpus are a shared influence on all of us, and thus the likelihood probabilities are reflective of the general correlations between emotions that we are routinely exposed to. As a result, the top experienced emotion chosen by the machine matches the top experienced emotion as chosen by the human evaluators as well. However, the distribution of possible experienced emotions was more spread out for the human evaluation. This could be due to the biases/perceptions of individual evaluators (evaluators evaluate the transcripts based on their personal experiences and prejudices). Moreover, some evaluators could have limited introspective abilities. Fatigue and lack of concentration in the task could be another reason for the flat distribution. 

This distribution of human evaluation thus contains valuable information. The mean of the distribution can inform us about the general beliefs (as is reflected by the top choices). Additionally, the tail end of the distribution could give us information about the beliefs and biases of individual turkers.

\section{Conclusions}
In this work, we investigated the effectiveness of some crowdsourcing methods in the highly subjective task of estimating experienced emotions of distressed users. The study revealed many interesting results.  First, we found that a simple voting consensus is as effective as an optimal aggregation method for the task considered. Second, we found that a machine learning algorithm that is not explicitly modeled to characterize evaluator biases is as reliable as the human evaluation in terms of assessing the most dominant experienced emotions. We believe a comparison of human and machine evaluation can also help in distinguishing aspects such as general beliefs and evaluator-specific beliefs. 
\vskip -0.1in

\bibliography{example_paper}
\bibliographystyle{icml2019}



\end{document}